\documentstyle[preprint,aps,epsf]{revtex}
\begin{document}
\draft
\sloppy
\scrollmode

\newcommand{\lfrac}[2]{#1/#2}
\renewcommand{\baselinestretch}{1.3}       %
\newcommand{\deltabar}{\,\,{\bar{}\hspace{1pt}\! \!\delta }}
\newcommand{\drbar}{\,\,{\bar{}\hspace{1pt}\! \!{\rm d} }}
\newcommand{\dbar}{\,\,{\bar{}\hspace{1pt}\! \!{d} }}

\title{New Action Principle for Classical
             Particle Trajectories in Spaces with Torsion}
\author{P.~Fiziev\thanks{Permanent address:
 Department of Theoretical Physics,
 Faculty of Physics, University of Sofia,
 Bull.~5 James Boucher, Sofia~1126,
 Bulgaria.
 Work supported by the Commission of the
 European Communities for Cooperation in Science
 and Technology, Contract~No. ERB3510PL920264
 }
 and H.~Kleinert}
\address{Institut f\"{u}r Theoretische Physik,\\
          Freie Universit\"{a}t Berlin\\
          Arnimallee 14, D - 14195 Berlin}
\date{\today}
\maketitle
\begin{abstract}
To comply with recent developments of path integrals
 in spaces with curvature and torsion we find
 the correct variational principle for the classical
trajectories. Although the action depends only on the length,
 the trajectories are {\em autoparallels\/} rather than geodesics
due to the effects of a new torsion force.
\end{abstract}
~\\~\\~\\
\newpage
\noindent
1)~Recently it was shown \cite{curv.a.tors,Pibuch}
   that in order to obtain a consistent path
  integral for point particles moving in a space with curvature
  and torsion, the classical trajectories must necessarily be
   autoparallels rather than geodesics.
Whereas geodesics are described by
\begin{equation}
  \ddot{q}^\nu + \bar\Gamma_{\alpha \beta }{}^\nu \dot{q}^\alpha
            \dot{q}^\beta  = 0,
\label{g}\end{equation}
with Christoffel symbols
$\bar{\Gamma }_{\mu \nu }{}^\lambda $,
the equations of
  motion of autoparallels are
\begin{equation}
 \ddot{q}^\nu  + \Gamma _{\alpha \beta }{}^\nu  \dot{q}^\alpha
          \dot{q}^\beta = 0,
\label{s}\end{equation}
where $\Gamma _{\mu \nu }{}^\lambda $ are the affine connections
(related to the Christoffel via
   $\Gamma _{\mu \nu \lambda}
  =\bar  \Gamma _{\mu \nu \lambda }
   + S _{\mu \nu \lambda }
   - S _{ \nu \lambda\mu }
   + S _{ \lambda\mu \nu }$
where $ S _{\mu \nu \lambda }$ is the torsion tensor).

This problem is quite general and appears in different branches of physics.
For example,
in the past, the general theory of relativity
has been extended to spaces with torsion \cite{grav}. Assuming
plausible actions and the usual field theoretic
formalism
it has been argued that the orbits of spinless particles
are geodesics. Since
up to now no possibility seems to exist to
falsify this result experimentally, the authors have been on safe
although rather academic grounds.
This situation has been changed drastically
with the findings in the context of path integrals
in Ref. [2]. It has been known for a long time that
the equations of motion of {\em nonrelativistic\/} 
celestial mechanics
can be solved most efficiently
by subjecting them to a
Kustaanheimo-Stiefel transformation \cite{kust}.
It has also been known that when transforming the action likewise,
its variation does {\em not\/} yield the correct equations of motion.
The reason lies in the
nonholonomic nature of the transformation.
This transformation
has been shown \cite{Pibuch} to
introduce torsion into the new coordinate system.
Thus, the description of celestial bodies
in these coordinates
can serve as an ideal experimentally well-known
testing ground
for any theory in spaces with torsion.
In Ref. [2] it was shown that the trajectories
of celestial bodies in this space
move along the autoparallels rather than the geodesics.
Although at first sight unrelated,
this undisputable fact calls for a revision of
a large body of theory developed in the context of
general relativity in spaces with torsion.

The fact that the
trajectories are autoparallels
agrees
with Einstein's equivalence principle according to which
the correct dynamical equations in a noneuclidean space
may be found by transforming the
 free-particle equations of motion in
a flat space $\ddot{x}^i(t) = 0$ locally to
 arbitrary curvilinear coordinates via
  some differential transformation
\begin{equation}
 dx^i = e^i{}_\mu (q)dq^\mu .
\label{1}\end{equation}
The transformed equations are postulated to
describe the orbits $q^ \mu (t)$
 in the curved space with the
metric 
\footnote{The new action principle 
we propose here 
has different physical applications.
In this article we ilustrate the main idea of this 
principle only on the simplest example 
of nonrelativistic particle in tree dimensional space.
Therefore we use here an euclidean metric in the x-space, 
which is enough for nonrelativistic problems. 
To get the relastvistic signature of the metric 
in the corresponding four-dimensional space with torsion
one has to put a  Minkovski metric into 
a flat four-dimensional  x-space.}
\begin{equation}
    g_{\mu \nu }(q) = e^i{}_\mu (q) e^i{}_\nu (q).
\label{m}\end{equation}
%
In the presence of torsion, the transformation (\ref{1})
 carries the equations of motion into
(\ref{s}), not (\ref{g}).

The key to the present work
lies in a problem which
 arises when trying
to apply
Einstein's
equivalence principle 
to the action
of the free particle
\begin{equation}
  {\cal A}= \int^{t_b}_{t_a} dt \frac{M }{2} \dot{x}^2(t).
\label{}\end{equation}
Under the transformation (\ref{1}), this goes
over into
\begin{equation}
  {\cal A} = \int^{t_b}_{t_a} dt \frac{M}{2} g_{\mu \nu }
       (q(t)) \dot{q}^\mu(t) \dot{q}^\nu (t)
\label{a}\end{equation}
If extremized in the standard way \cite{3}, it would yield
the euqations
of motion (\ref{g}),
i.e., the
 the wrong trajectories in the presence
 of torsion.

The purpose of this note is to resolve
this conflict by
exhibiting a new variational principle in spaces
with torsion  where action (\ref{a}) yields
 autoparallels after all.

~\\
\noindent
2)~The important point to realize is that in order
  to obtain the correct trajectories from the transformed action
 (\ref{a}), the variations
 $\delta q^\mu (t)$ have to be performed
in accordance with {\em independent\/}
  variations $\delta x^i(t)$ in the flat space
 which vanish at the endpoints (see Fig.~1a).
 The paths in the two spaces
   are related by the integral equation
\begin{equation}
    q^\mu (t) = q^\mu (t_a) + \int^{t}_{t_a}
       e_i{}^\mu (q(t')) \dot{x}^i(t') dt'
\label{r2}\end{equation}
where $e_i{}^\mu (q)$ is the inverse matrix of
 $e^i{}_\mu (q)$ in (\ref{1}).
It is easy to derive from this an explicit, albeit nonlocal,
equation between $\delta x^i(t)$
 and $\delta q^\mu (t)$. For this we introduce
the auxiliary quantities
\begin{equation}
  \delta _hq^\mu  \equiv e_i{}^\mu (q) \delta x^i
\label{}\end{equation}
which are equivalent to the
independent variations $\delta x^i(t)$
to be performed
 on the system.
From (\ref{1})  we find the relation
\begin{equation}
    \delta \dot{q}^\mu (t) = \frac{d}{dt}
           \delta _h q^\mu +2 \dot{q}^\nu  S_{\nu \lambda} {}^\mu
            \delta _nq^\lambda -\delta \Delta ^\lambda
             \Gamma _{\lambda \nu }{}^\mu \dot{q}^\nu
\label{3}\end{equation}
where
\begin{equation}
   \delta \Delta ^\mu \equiv \delta q^\mu -\delta _hq^\mu.
\label{}\end{equation}
Using (\ref{r2}) we see that
\begin{equation}
 \delta \dot{q}^\mu (t) = \frac{d}{dt} \delta q^\mu (t)
\label{}\end{equation}
 so that (\ref{3}) can be rewritten as
\begin{equation}
   \frac{d}{dt} \delta \Delta = - \delta \Delta ^\lambda
            \Gamma ^\mu _{\lambda \nu } \dot{q}^\nu
             + 2 \dot{q}^\nu S _{\nu \lambda }{}^\mu
              \delta _h q^\lambda .
\label{}\end{equation}
This differential equation can be integrated to
\begin{eqnarray}
   \delta \Delta (t) = \int^{t}_{t_a} dt'
          U_{tt'}~~\sigma (t')
\label{equ}\end{eqnarray}
where
\begin{equation}
  U_{tt'} = T \exp \left[ -\int^{t}_{t'} G(t'')dt''\right].
\label{}\end{equation}
We have used vector and matrix notation with
\begin{eqnarray}
 \label{15} G_\mu {}^\lambda (t) & = & \Gamma _{\mu \nu }{}^\lambda
               (q(t))\dot{q}^\nu (t)\\
   \sigma ^\mu (t) & = & 2 \dot{q}^\nu (t)  S_{\nu\lambda} {}^\mu
                 (q(t))
                \delta _hq^\lambda (t) .
\label{}\end{eqnarray}
If the space is free of torsion, then $\sigma (t)
 \equiv 0$ and hence $\delta \Delta (t) \equiv 0$. In this case
the variations $\delta q^\mu (t)$ coincide with
 the independent ones $\delta _h q^\mu (t)$.
We shall therefore refer to $\delta _h q^\mu (t)$ as
 the {\em holonomic\/} variations of the
  path $q^\mu (t)$.
In the presence of torsion, on the other hand, the variations
 $\delta q^\mu (t)$ differ from $\delta _h q^\mu (t)$.
We shall emphasize the nonholonomic properties of $ \delta q^ \mu $
by writing them as  $ \deltabar q^ \mu $, in analogy with the
notation   $ \drbar Q $  used in thermodynamics
for a small increment of a nonintegrable function.

Let us calculate the variation of the
action which is
\begin{equation}
  \deltabar A = M\int^{t_b}_{t_a} dt \left( g_{\mu \nu }
               \dot{q}^\nu \deltabar \dot{q}^\mu + \frac{1}{2}
               \deltabar q^\mu \partial _\mu g_{\alpha \beta }
                \dot{q}^\alpha \dot{q}^\beta \right)
\label{var}\end{equation}
The most direct way of deriving
from $ \deltabar A$ the equations of motion is by inserting
an infinitesimal local ``knock'' variation
\begin{equation}
 \delta _hq^\mu (t) = \epsilon ^\mu \delta (t-t_0).
 \label{17}\end{equation}
In the torsion-free case with $\deltabar q^\mu (t) \equiv \delta _h
  q^\mu (t)$
we obtain
\begin{equation}
   \deltabar A\equiv \delta _h A = - \epsilon ^\mu
        Mg_{\mu \nu }(\ddot{q}^\nu + \bar{\Gamma }_{\alpha \beta }{}^\nu
        \dot{q}^\alpha \dot{q}^\beta ).
\label{}\end{equation}
Setting this equal to zero produces at once
the equations of motion which are those of
 geodesics, i.e., the correct particle
 trajectories in the absence of torsion.

In the presence of torsion, the holonomic ``knock''
  variation (18)
   leads via (13) - (16) to the
total variation
\begin{equation}
   \deltabar q(t) = \epsilon \delta (t-t_0) + \deltabar \Delta (t)
\label{}\end{equation}
with
\begin{equation}
  \deltabar \Delta (t) = \Theta (t-t_0)U_{tt_0}\sigma (t_0)
\label{}\end{equation}
Note that the term $\deltabar \Delta (t)$ is nonlocal.
It has the important effect that
while $\deltabar q^\mu (t_a)=0$,
 the variation at the endpoint $\deltabar q^\mu (t_b) = b^\mu $
 is nonzero.
 The vector $b^\mu $ corresponds to the Burgers vector
 in the physics of dislocations [4]. The situation
 is illustrated in Fig.~1c.

The nonlocal term modifies the usual
 derivation of the equations of motion.
The time derivative of $\deltabar \dot{q}(t)$ is
\begin{equation}
  \deltabar \dot{q}(t) = \epsilon \dot{\delta }(t-t_0)
                  + \delta (t-t_0) \Sigma (t_0)\epsilon -
           G\deltabar \Delta (t)
\label{q.}\end{equation}
where $\Sigma_\mu {}^\lambda (t) = 2 S_{\mu \nu }{}^\lambda
           (q(t)\dot{q}^\nu (t)$.
Inserting $\deltabar q^\mu (t)$ and
 $\deltabar \dot{q}^\mu (t)$ into (17)
we obtain $\deltabar A = \delta _hA + \deltabar _aA$, with
 the anholonomic deviation $\deltabar _aA$ of the action
\begin{equation}
 \deltabar _a A= - \epsilon ^\mu 2MS_{\mu \alpha \beta }(q(t_0))
      \dot{q}^\alpha (t_0) \dot{q}^\beta (t_0)
\label{}\end{equation}
originating from the second term in (22).
The contribution of $ -G \deltabar \Delta (t)$ in (22)
cancels exactly the contribution of the term
 $\deltabar \Delta ^\mu \frac{1}{2} \partial _\mu g_{\alpha \beta }
  \dot{q}^\alpha \dot{q}^\beta $ which originates from the
second term in $\deltabar A$ (17), because of
(15) and the identity
$\frac{1}{2} \partial _\mu g_{\alpha \beta } -\Gamma _{\mu \{
\alpha \beta \}}\equiv 0$ which follows directly from the definitions
$g_{\alpha \beta }= e^i{}_\alpha e^i{}_\beta $ and
 $\Gamma _{\mu \nu }{}^\lambda  = e_i{}^\lambda
 \partial _\mu e^i{}_\nu $ (it expresses the fact that the covariant
derivative of the metric tensor vanishes, $D_ \mu g_{\alpha \beta } \equiv 0$.).

Hence the total variation of the action is
\begin{eqnarray}
  \deltabar A & = & -M\epsilon ^\mu g_{\mu \nu }\left[ \ddot{q}^\nu +
     \left( \bar{\Gamma }_{\alpha \beta }{}^\nu -2S^\nu _{\alpha \beta }
 \right) \dot{q}^\alpha \dot{q}^\beta \right] \nonumber \\
   & = & -M\epsilon ^\mu g_{\mu n}\left( \ddot{q}^\nu  +
          \Gamma _{\alpha \beta }{}^\nu \dot{q}^\alpha
           \dot{q}^\beta \right) ,
\label{}\end{eqnarray}
the second line following from the first via the identity
$\bar{\Gamma }_{\alpha \beta }{}^\nu  = \Gamma _{\{\alpha \beta \}}{}^\nu
 + 2 S^\nu _{\{\alpha \beta \}}$.
Setting $\deltabar A=0$ gives the autoparallels as
the equations of motions
which is what we wanted to demonstrate.

~~{}\\[5mm]
\newpage
~\\
{
{\bf FIGURE 1}
~~~In the usual holonomic case, the paths $x(t)$ and $x(t)+\delta x(t)$
shown in (a)
are mapped into
paths $q(t)$ and $q(t) + \delta_h q(t)$
shown in (b).
In the
nonholonomic case with $S^\nu_{\alpha \beta } \neq 0$, however,
they go over into
$q(t)$ and $q(t)+\deltabar q(t)$
shown in (b)
with a
failure to close at $t_b$ by a vector ${\bf b}$, the analog
 to the Burgers vector in a solid with dislocations. \label{Fig.1}
}
%
%

\end{document}